\input epsf.tex
\def\DESepsf(#1 width #2){\epsfxsize=#2 \epsfbox{#1}}
\documentclass{ws-procs9x6}

\begin{document}

\title{ Cosmology In Horava-Witten M-Theory
}

\author{R. Arnowitt, James Dent and B. Dutta\footnote{Present address: Department of Physics, University of Regina, Regina
SK, S4S  0A2 Canada}}

\address{Center For Theoretical Physics, Department of Physics, Texas A$\&$M University,
College Station TX 77843-4242, USA\\ 
E-mail: arnowitt@physics.tamu.edu, j-dent@physics.tamu.edu, duttabh@uregina.ca}


\maketitle

\abstracts{The cosmology of the Horava-Witten M-Theory reduced to five dimensions
retaining the volume modulus is considered. Brane matter is considered as a
perturbation on the vacuum solution, and the question of under what
circumstances does the theory give rise to the standard RWF cosmology is
examined. It is found that for static solutions, one obtains a consistent
solution of the bulk field equations and the brane boundary conditions only
for pure radiation on the branes. (A similar result holds if additional
5-branes are added in the bulk.) If one stabilizes the fifth dimension in
an ad hoc manner, a similar inconsistency still occurs (at least for a
Hubble constant that has no dependence on y, the fifth dimension.) Within
this framework, the possibility of recovering the RWF cosmology still remains
if the volume modulus and /or the distance between branes becomes time
dependent, under which circumstances the Hubble constant must then depend on y
(unless the fifth dimension and volume modulus expand at precisely the same
rate).
}

The problem of obtaining a satisfactory cosmology in 5 dimensional (5D)
theories so that the square of the Hubble constant H is proportional to the
mass density has been the subject of much discussion since the initial work
of Binetruy, Deffayet and Langlois\cite{bdl}. Thus in the Randall-Sundrum 1 (RS)
type model\cite{rs} with
two  branes separated in the fifth dimension, it has been argued that it is
insufficient to assume a static solution, and it is necessary to
dynamically stabilize the fifth dimension to obtain the standard
Robertson-Walker-Friedman (RWF) cosmology\cite{cgrt} (see also talk by P.Kanti,
these proceedings), though alternate views have been expressed\cite{kz}. The
Horava-Witten (H-W) 11D theory\cite{hw}, compactified on a Calabi-Yau (CY)
manifold becomes a 5D theory at low energy\cite{losw}. Here it has been argued
in Refs.\cite{e,ekr}
 that
the correct RWF cosmology does arise (without stabilization) in the Hubble
expansion era. However, in both these analyses, the boundary
conditions on both end branes have not been imposed (and in Ref.\cite{ekr}, the
equations in the bulk have not been solved to check consistency). An
approximate solution in the bulk satisfying the brane boundary conditions
for the inflationary era was obtained in\cite{low}, but this solution does not
apply to Hubble expansion era where the string  constants are much larger
than the brane matter densities.

In this discussion we will consider the H-W theory reduced to five
dimensions for the simple case where we retain a volume moduli V. (Shape
moduli will be neglected). We will solve the bulk equations for the static
solution, impose fully all the boundary conditions on the two branes, and
examine under what circumstances one can generate the conventional 4D RWF
cosmology. We will see that the existence of moduli distinguishes the H-W
M-theory from the phenomenological RS model.

\section{ Horava-Witten M-Theory} In this section we briefly summarize the Horava-Witten theory which is 11D
supergravity on an orbifold: $M_{10}\times X\times S^1/Z_2$ where $Z_2$ is the parity
operation on the eleventh dimension. Yang Mills multiplets with gauge group
$E_8$ exist on the orbifold planes at the fixed points ($x^{11}$ = 0 and $\pi
\rho$)  to cancel gravitational anomalies, and then cancellation  of the 10D
fermion anomaly produces a relation bewteen the Yang Mills gauge coupling
constant and the 11D gravitational constant\cite{hw}. This resolves naturally
the puzzle of conventional 10D heterotic string theory of why grand
unification occurs at the GUT scale $M_G \simeq3\times10^{16}$ GeV, rather than at
the 4D Planck mass, for it is the 11D Planck constant that is fundamental,
and its mass scale is $\sim M_G$ (while the orbifold length is $\sim 1/(5\times10^{15}$GeV)
i.e. about 10 times the compactifcation scale).  If we retain only the
Calabi-Yau volume modulus $V$, the reduced 5D theory Lagrangian takes the
following simple form\cite{losw}:
\begin{eqnarray}
L&=&-{1\over{2 k_5^2}}\int_{M_5}\sqrt{-g}[R+{1\over
2}V^{-2}\partial_{\alpha}V\partial^{\alpha}V+{3\over 2}\alpha^2]\\\nonumber&+&{1\over
k_5^2}\Sigma_i\int_{M_4^{i}}\sqrt{-g}V^{-1}3\alpha(-1)^{i+1}\alpha
-{1\over {16\pi\alpha_G}}
\Sigma_i\int_{M_4^{i}}\sqrt{-g}Vtr{F^{(i)}}^2_{\mu\nu}\\\nonumber
&-&\Sigma_i\int_{M_4^{i}}\sqrt{-g}[(D_{\mu}C)^n(D_{\mu}{\bar
C})^n+V^{-1}{{\partial W}\over{\partial C^n}}{{\partial {\bar W}}\over{\partial
{\bar C}^n}}+D^{u}D^{u}]\end{eqnarray}
Note that the bulk cosmological constant and the brane cosmological
constants are fixed by the same parameter $\alpha$ (which is scaled by the
(integer) first Pontryagin class of the CY manifold). Thus unlike the RS
model, there is no freedom to fine tune the cosmological constants.
Numerically $\alpha=O(10^{15}\rm GeV)$ and so is much larger than the matter
density (except in the inflationary era considered in\cite{low} where they are comparable).

We chose the following form for the 5D metric to describe he Hubble
expansion era:
\begin{equation}
                                        ds^2 = a^2 dx^kdx^k - n^2 dt^2 +
b^2 dy^2  \end{equation}
where $y = x^{11}$ and a, n and b are in general functions of y and t. The
relevant field equations are the Einstein equation arising from $G_{tt}$,
$G_{kk}$,
$G_{yy}$, $G_{ty}$ and the equation for V, and are given in e. g.\cite{ekr}. In addition
there are the boundary conditions on the 3 branes at the orbifold planes $y_1
= 0$ and $y_2 = \pi \rho$:

\begin{eqnarray}
(-1)^i({1\over b}{a'\over a})_{y_i}&=&{1\over{6M^3}}\rho_i\\\nonumber
(-1)^i({1\over b}{n'\over n})_{y_i}&=&-{1\over{6M^3}}(2\rho_i+3 p_i)\\\nonumber
(\phi')_{y_i}&=&[(3b\alpha-{b\over M^3}\rho_{inr})e^{-\phi}]_{y_i}
\end{eqnarray}
where $M$ is the 5D Planck mass, $\rho_i$ and $p_i$ are the total matter density and
pressure on the two 3 branes, $V = e^{\phi}$, and $\rho_{i(nr)}$ is the
massive (non-relativistic) matter. (Note that Ref.\cite{ekr} omits the effects of
the $V$ factors in this analysis.)
\section{Solution Of The 5D Equations}

The cosmological constants in the bulk and on the banes are of GUT size,
while the matter density coming from the branes are generally very small
throughout the Hubble expansion era. Thus one can solve the field equations
perturbatively, first neglecting the brane matter, and then including it in
higher approximations. This should be a good approximation except in the
very early universe during the inflation era where there $(\rho) \sim (M_G)^4$.
To zero'th approximation then, we neglect all brane matter which leads to
the static vacuum solution of Ref.\cite{losw}: $a(y) = n(y) = f^{1/2}$; $b(y) = b_0
f^2$; $V(y) = b_0 f^3$, where $f(y) = c_0 + \alpha |y|$. This solution solves all
field equations and boundary conditions exactly, preserves Poincare
invariance and breaks 4 of the 8 supersymmetries, which is appropriate for
getting N=1 supergravity when one descends to four dimensions. Thus it
represents an appropriate choice for a vacuum metric. The parameters $a_0$,
$b_0$, $c_0$ are arbitrary due to the existence of flat directions in the
potential (which perhaps would be determined by non-perturbative
contributions). We next include the matter on the brane as a perturbation
that generates the Hubble expansion and calculate it's effects to linear
order in $\rho$. We consider first a solution where both the fifth dimension
and the CY volume remains static, and only $a(y,t)$ has time dependence to
accommodate Hubble expansion:
\begin{eqnarray}
a(y,t) &=& f^{1/2} (1 + \delta a(y,t)); \, n(y) = b_0f^{1/2} (1 + \delta
n(y));\\\nonumber   b(y)&=& b_0f^2 (1 + \delta b(y));\, V(y) = b_0 f^3(1 + \delta V(y))
\end{eqnarray}
The Hubble constant is then  $H = {\dot{a}}/a \simeq {\delta \dot{a}}$, where `` (dot, prime)"
means ``$(t,y)$" derivative. Similarly one has $\delta {\ddot{a}} =
{\dot{H}}  + H^2$.
The boundary conditions of Eq. (3) then reduce to ($i = 1,2$ for the two branes):
\begin{eqnarray}
(\delta a'+{1\over 2}{\alpha\over f}\delta
V)_{i}&=&{(-1)^i\over{6M^3}}({\rho\over f})_i\\\nonumber
(\delta n'+{1\over 2}{\alpha\over f}\delta V)_{i}&=&-{(-1)^i\over{6M^3}}({{2\rho+3
p}\over f})_i\\\nonumber
(\delta V'+{3}{\alpha\over f}\delta V)_{i}&=&-{(-1)^i\over{M^3}}({{\rho_{nr}}\over f})_i
\end{eqnarray}
For the static solution, one may chose a special frame where $\delta b = 0$.
At first sight then, it would appear that a satisfactory solution should be
available: there are three independent variables, $\delta a$, $\delta n$, $\delta
V$
obeying second order differential equations, and so  there will be a total
of six constants of integrations which in addition to H would give seven
quantities to satisfy the six boundary conditions of Eqs. 5. However, the
boundary conditions are two point conditions (rather than ``initial"
conditions), and as is well known such conditions need not give consistent
solutions to differential equations (e.g. a free particle where we specify
the initial and final velocity (rather than the initial position and
initial velocity) generally has no consistent solution without ``fine
tuning" the velocity choice). Indeed, the equations from $G_{tt}$, $G_{kk}$,
$G_{yy}$
and $V$ allow one to obtain a first integral involving precisely the
quantities appearing in Eq. (5):

\begin{eqnarray}
(\delta a'+{1\over 2}{\alpha\over f}\delta
V)_{i}&=&{b_0^2\over\alpha}[\int^f_{f_1}df f^3H^2+C_1]\\\nonumber
(\delta n'+{1\over 2}{\alpha\over f}\delta
V)_{i}&=&{b_0^2\over\alpha}[\int^f_{f_1}df f^3[H^2+2{\dot{H}}]+C_2]\\\nonumber
(\delta V'+{3}{\alpha\over f}\delta V)_{i}&=&-3{b_0^2\over\alpha}[\int^f_{f_1}df
f^3[4H^2+2{\dot{H}}]+C_3]
\end{eqnarray}

In Eqs.(6), we have left the y dependence of $H (y,t)$ arbitrary though the
simplest choice (e.g. used in \cite{cgrt}) would be to assume $H$ to be  independent of y.
In spite of this we see that there are only three constants of integration
to satisfy the six constraints of Eqs.(6), and so a solution will exist
only under special situations.

The boundary conditions of the first two of Eqs.(5) then yield the following results:
\begin{eqnarray}
\int^{f_2}_{f_1}f^3H^2&=&{1\over 6}\lambda({\rho_{2}\over f_2}+{\rho_{1}\over
f_1});\,C_1=-{\lambda\over
6}{\rho_{i}\over f_i}\\\nonumber
\int^{f_2}_{f_1}f^3{\dot H}&=&{1\over 4}\lambda({\rho_{2}+p_2\over
f_2}+{\rho_{1}+p_1\over
f_1});\,C_2=-{\lambda\over
2}{p_{i}\over f_i}\\\nonumber
\end{eqnarray}
where $\lambda = \alpha/ b_0^2 M^3$. One has
\begin{equation}{\dot{\rho}} = - 3 H (\rho +p)
\end{equation}
as a consequence of the Bianchi identities for the static solution,
or alternately from the $G_{ty}$ equation\cite{ekr}. If $H$ is independent of $y$, one may
integrate Eq.(7) and obtain the conventional relation between $H^2$ and $\rho$
(with an explicit evaluation of the 4D Newtonian constant $G_N$). However, the
third  equation in Eq.(5) imposes an additional constraint yielding:
\begin{equation}
\int^{f_2}_{f_1}f^3[4H^2+2{\dot H}]=
{1\over 3}\lambda({\rho_{2(nr)}\over f_2}+{\rho_{1(nr)}\over f_1});\,C_3={1\over
3}\lambda{\rho_{i(nr)}\over f_i}
\end{equation}
Inserting Eqs.(7) into Eq.(9) then requires
\begin{equation}
{{\rho_2-3 p_2}\over f_2}+{{\rho_1-3 p_1}\over f_1}=2({\rho_{2(nr)}\over f_2}+{\rho_{1(nr)}\over f_1})
  \end{equation}
which requires $p_i = 1/3\rho_i$, and $\rho_{i(nr)} = 0$. Thus a consistent
solution is obtained only for pure radiation, and non-relativistic matter
(or a cosmological constant) is not accommodated by a static solution in
the Horava-Witten model. This is true independent of the y dependence of $H$
\section{Alternate Possibilities}

One may extend the results of Sec. 3 to include an arbitrary number of
5-branes in the bulk\cite{cgls}. One again finds (at least for a Hubble constant
that is independent of y) that a consistent solution occurs only if pure
radiation is on the orbifold branes. The RS model is a phenomenology, and
so one may add additional fields into the bulk in an ad hoc manner. In this
way the fifth dimension can be stabilized by including a potential for a
radion field. The RS model then gives rise to a correct RWF cosmology for
arbitrary matter. In contrast, the H-W M-theory is a reasonably well defined
theory, and one is not free to add additional fields into the bulk at will.
However, one might speculate on the possibility that some non-perturbative
potential exists between the two orbifold 3-branes that stabilizes the
fifth dimension. This would correspond to discarding the $G_{yy}$ field
equation\cite{cgrt}. One may still solve the remaining field equations for the static
case, and again one finds that for a y-independent Hubble constant, the
equations are inconsistent (except for the radiation case). A linear
y-dependence also is inconsistent. (Of course if by hand one also discards
the V modulus equation as well, one can get consistent solutions as the
theory is then reduced to the RS model.)

The remaining possibility then is to consider dynamic solutions where the
volume modulus becomes time dependent. This possibility is not without some
interest, as it would imply that the gauge coupling constant would be time
dependent over cosmological time scales, for which some data exists\cite{w}.
In this circumstances, one may show that the Hubble constant then must have
y-dependence, the Hubble constant on the distant brane being different from
the one on the physical brane (unless the fifth dimension expands at precisely
the same rate as the volume modulus). This more complicated possibility is under
investigation.

\section{Acknowledgement} This work was supported in part by the National Science
Foundation Grant PHY-0101015.

\end{document}